\documentclass[a4paper]{jpconf}
\usepackage{graphicx}
\usepackage[hidelinks]{hyperref}
\usepackage{color}
\usepackage{amsmath,amsfonts,amssymb,bm}

\usepackage{lipsum}

\usepackage{fancyhdr}
\def\beq{\begin{equation}}
\def\eeq{\end{equation}}
\def\beqn{\begin{eqnarray}}
\def\eeqn{\end{eqnarray}}

\def\r {{\bf r}}

\def\P {{\bf P}}

\def\bi {\mbox{\boldmath $i$}}

\def\r {{\bf r}}

\def\P {{\bf P}}

\begin{document}

\title{Variance of a Trapped Bose-Einstein Condensate}

\author{O~E~Alon$^{1,2}$}
\address{$^{1}$ Department of Mathematics, University of Haifa, Haifa, Israel}
\address{$^{2}$ Haifa Research Center for Theoretical Physics and Astrophysics, University of Haifa, Haifa, Israel}

\ead{ofir@research.haifa.ac.il}

\begin{abstract}
The ground state of a Bose-Einstein condensate in a two-dimensional trap potential
is analyzed numerically at the infinite-particle limit.
It is shown that the anisotropy of the many-particle position variance along the $x$ and $y$
axes can be opposite when computed at the many-body and mean-field levels of theory. 
This is despite the system being $100\%$ condensed,
and the respective energies per particle and densities per particle to coincide.
 \end{abstract}

\section{Introduction}\label{Intro}

We consider the ground state of $N$
interacting bosons in a two-dimensional trap.
It has been shown under quite general conditions 
\cite{INF1,INF2,INF3} 
that the many-body energy per particle and density per particle coincide at the infinite-particle limit
with the Gross-Pitaevskii mean-field results,
\beq\label{E1}
 \lim_{N \to \infty} \frac{\rho(\r)}{N} = |\phi_{GP}(\r)|^2, \qquad
 \lim_{N \to \infty} \frac{E}{N} = \varepsilon_{GP}.
\eeq
Here, the density is the diagonal of the reduced one-particle density matrix \cite{DENS1,DENS2},
$\rho(\r) \equiv \rho^{(1)}(\r,\r)$,
$E$ is the ground-state energy,
$\phi_{GP}(\r)$ and $\varepsilon_{GP}$ are the Gross-Pitaevskii orbital and energy, respectively,
and $\r=(x,y)$.
Furthermore, the bosons are $100\%$ condensed,
at the levels of the reduced one-particle and two-particle (and any finite-order \cite{DENS3}) density matrices,
\beqn\label{E2}
& & \lim_{N \to \infty} \frac{\rho^{(1)}(\r_1,\r_1')}{N} = \phi_{GP}(\r_1)\phi^\ast_{GP}(\r_1'), \nonumber \\
& & \lim_{N \to \infty} \frac{\rho^{(2)}(\r_1,\r_2,\r'_1,\r'_2)}{N(N-1)} = 
\phi_{GP}(\r_1)\phi_{GP}(\r_2)\phi^\ast_{GP}(\r'_1)\phi^\ast_{GP}(\r'_2). \
\eeqn
The infinite-particle limit implies that the interaction parameter, i.e., the product of the number of particles times the interaction strength, is kept fixed when $N \to \infty$.
The question then arises, which differences are there, at the infinite-particle limit,
between the many-body and mean-field descriptions of trapped bosons in their {\it ground state}.

To research this topic, it has been pointed out recently that the variance of many-particle
operators can have substantial differences when computed at the many-body and mean-field levels of theory,
even when the bosons are $100\%$ condensed \cite{VAR1}.
In particular, the variance of the many-particle position operator, $\hat X = \sum_{j=1}^N \hat x_j$,
\beq\label{E3}
\frac{1}{N} \Delta^2_{\hat X} = \int d\r \frac{\rho(\r)}{N}x^2 - N\left[\int d\r \frac{\rho(\r)}{N}x\right]^2
+ \int d\r_1 d\r_2 \frac{\rho^{(2)}(\r_1,\r_2,\r_1,\r_2)}{N}x_1x_2,
\eeq
picks up tiny fluctuations that are completely washed out in the above-mentioned
properties at the infinite-particle limit.
This is because the two-particle reduced density matrix in the last term of (\ref{E3})
is divided only by $N$ and not by $N(N-1)$ as in (\ref{E2}).
Another property of interest is the overlap between the many-body and mean-field wavefunctions,
which is always less than and can become much smaller than $1$ \cite{INF3,INF4}.

\begin{figure}[!]
\begin{center}
\includegraphics[width=0.645\columnwidth,angle=-90]{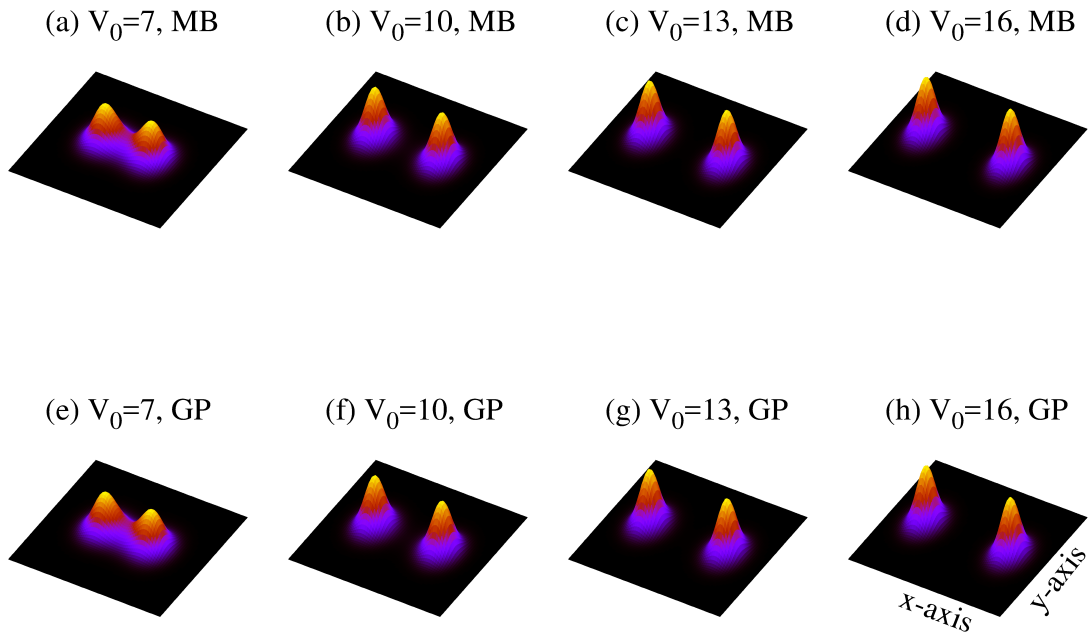}
\end{center}
\caption{Density per particle of the ground state of $N=1,000,000$ bosons
in two-dimensional double-well traps for four barrier heights $V_0$.
Top row, panels (a)-(d): Many-body computations with $M=2$ self-consistent orbitals.
Bottom row, panels (e)-(h): Mean-field computations with $M=1$ self-consistent orbitals, i.e., Gross-Pitaevskii results.
The respective densities per particle are indistinguishable.
On the other hand,
the anisotropies of the many-body and mean-field position variances along the $x$ and $y$ directions 
become opposite with increasing barrier height.
See Fig.~\ref{f4} and the text for further details.
The quantities shown are dimensionless.}
\label{f1}
\end{figure}

Generally, for repulsive interactions the many-body position variance is smaller than the mean-field variance,
and vice versa for attractive interactions.
This result is transparent to see analytically when considering bosons in an harmonic trap,
and requires accurate numerics to arrive at for anharmonic traps \cite{VAR1,VAR2}.
In two spatial dimensions,
one is drawn to look for properties and phenomena that cannot take place in one dimension.
The anisotropy of the position and momentum variance along the $x$ and $y$ directions 
was investigated in \cite{VAR3} and the variance of the many-particle
angular-momentum operator in \cite{VAR4}.
So far, it has not been shown that the ground state can exhibit opposite anisotropy of
the variance at the infinite-particle limit,
i.e., that the many-body and mean-field anisotropies of $100\%$ condensed bosons are different.
This is the main finding of the present work,
i.e., that trapped Bose-Einstein condensates at the infinite-particle limit 
can be classified according to
whether the anisotropies of their many-body and mean-field position variances 
are alike or opposite.

\section{System and results}\label{Result}

We consider $N$ repulsively interacting bosons in a two-dimensional double-well potential.\break\hfill
The height of the barrier is varied and we follow the changes in the system.
The interaction parameter $\Lambda=\lambda_0(N-1)$ is held fixed and the number of bosons is increased
towards the infinite-particle limit.
We see saturation of the quantities under investigation with $N$ for a given barrier height,
thereby providing strong numerical support of the conclusions to be made at the infinite-particle limit.

\begin{figure}[!]
\begin{center}
\includegraphics[width=0.545\columnwidth,angle=-90]{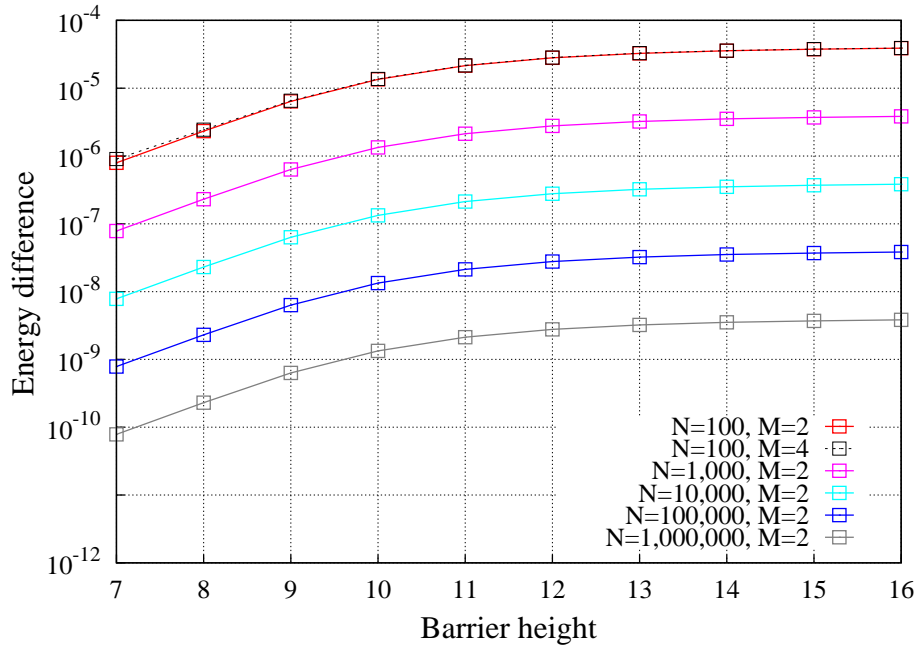}
\end{center}
\vglue 0.5 truecm
\caption{Difference between the mean-field
and many-body energies per particle for $N=10^2,\ldots,10^6$ bosons
as a function of the barrier height.
The interaction parameter is $\Lambda=\lambda_0(N-1)=0.1$.
The difference decreases with increasing number of particles $N$ for a given barrier height $V_0$.
See the text for further details.
Actual data is marked by symbols, the continuous curves are to guide the eye only.
The quantities shown are dimensionless.
}
\label{f2}
\end{figure}

The many-particle Hamiltonian whose ground state we are to investigate
is $\hat H(\r_1,\ldots,\r_N) = \sum_{j=1}^N \hat h(\r_j) + \lambda_0\hat W(\r_j-\r_k)$.
The one-particle Hamiltonian is $\hat h(\r) = -\frac{1}{2}\left(\frac{d^2}{dx^2} + \frac{d^2}{dy^2}\right) + 
\frac{1}{2} x^2 + \frac{1}{4}y^2 + V_0 e^{-\frac{x^2}{8}}$.
Note that the harmonic part of the confining potential is anisotropic,
and wider along the $y$ direction than along the $x$ direction.
The inter-particle interaction is Gaussian,
$\lambda_0 \hat W(\r) = \frac{\lambda_0}{2\pi\sigma^2}e^{-\frac{x^2+y^2}{2\pi\sigma^2}}$,
with width of $\sigma=0.25$.
Throughout this work $\Lambda=0.1$.

The ground-state of the trapped bosons is computed within the
multiconfigurational time-dependent Hartree for bosons (MCTDHB) method \cite{MCTDHB1,MCTDHB2}.
We use the numerical implementation in \cite{Package1,Package2}.
The method is well documented, used, benchmarked, and extended in the literature 
\cite{LT1,LT2,LT3,LT4,LT5,LT6,LT7,LT8,LT9,LT10,LT11,LT12,LT13,LT14,LT15,LT16,LT17},
and a brief account suffices here.
In the MCTDHB method, the self-consistent ground state \cite{MCHB} is obtained by imaginary-time propagation
and determined according to the variational principle.
The wavefunction is expanded by all time-dependent permanents, where $N$ bosons are distributed over 
$M$ optimized orthonormal one-particle functions,
with time-dependent coefficients,
such that the energy of the ground state is minimized.
For $M=1$ orbitals MCTDHB boils down to Gross-Pitaevskii theory.

The Hamiltonian is represented by an equidistant grid of $128 \times 128$ points
in a box of size $[-10,10) \times [-10,10)$ with periodic boundary conditions.
Convergence with respect to the grid size has been
verified using a $256 \times 256$ points
for the highest barrier, $V_0$, and largest number of self-consistent orbitals, $M$, employed.

\begin{figure}[!]
\begin{center}
\includegraphics[width=0.545\columnwidth,angle=-90]{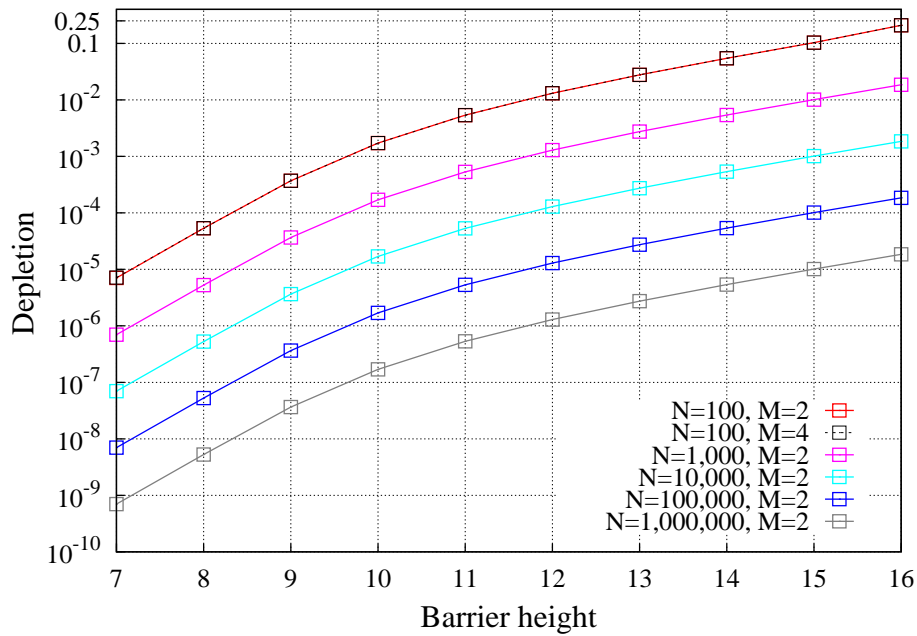}
\end{center}
\vglue 0.5 truecm
\caption{Depletion per particle $1-\frac{n_1}{N}$ 
for $N=10^2,\ldots,10^6$ bosons as a function of the barrier height.
The interaction parameter is $\Lambda=\lambda_0(N-1)=0.1$.
The depletion per particle decreases with increasing number of particles $N$ for a given barrier height $V_0$.
See the text for further details.
Actual data is marked by symbols, the continuous curves are to guide the eye only.
The quantities shown are dimensionless.
}
\label{f3}
\end{figure}

Fig.~\ref{f1} displays the densities per particle of $N=1,000,000$ bosons
computed at the $M=2$ many-body and $M=1$ mean-field levels of theory.
The respective densities per particle cannot be distinguished from each other,
in accordance with the left-hand-side of (\ref{E1}).
The accuracy and adequacy of the $M=2$ computations for the
two-dimensional trap and various barrier heights considered here
are verified against $M=4$ computations,
see below. 

We now examine other quantities and their values as the infinite-particle limit is taken.
Fig.~\ref{f2} depicts the difference between the mean-field and many-body
ground-state energies per particle as a function of the barrier height $V_0=7,\ldots,16$ 
and the number of particles $N=10^2,\ldots,10^6$, all at a constant interaction parameter $\Lambda$.
The difference is found to increase with $V_0$ for a given $N$
and to decrease with $N$ for a given $V_0$.
The latter is in accordance with the right-hand-side of (\ref{E1}).
To verify that $M=2$ self-consistent orbitals accurately describe the energy
in the two-dimensional double well for the various barrier heights,
we have performed calculations using $M=4$ and $N=100$ bosons,
which are seen to fall on top of the $M=2$ and $N=100$ bosons results,
see Fig.~\ref{f2}.
Since the interaction parameter is kept fixed,
this implies the accuracy of the $M=2$ and $N>100$ results as well.

\begin{figure}[!]
\begin{center}
\includegraphics[width=0.545\columnwidth,angle=-90]{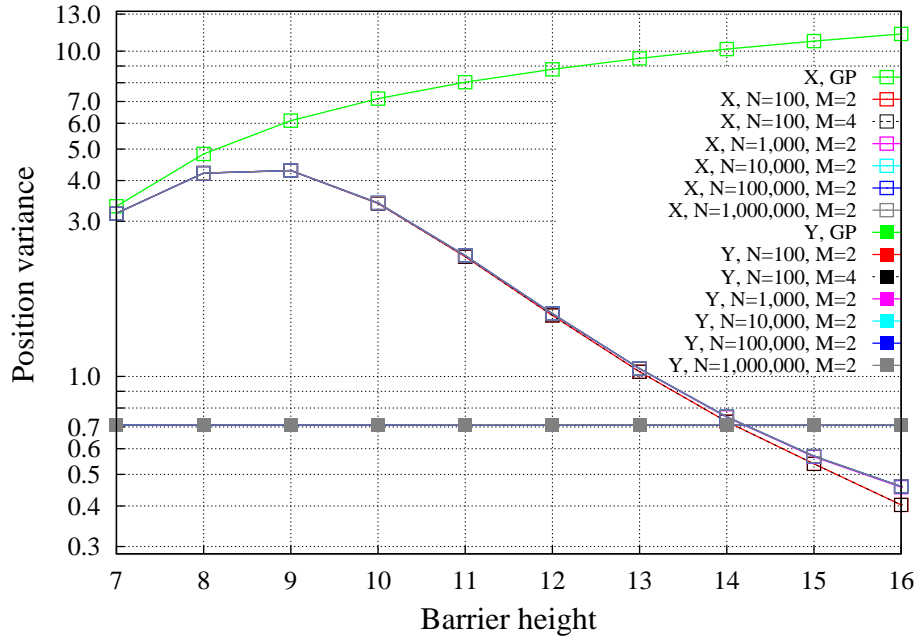}
\end{center}
\vglue 0.5 truecm
\caption{Many-particle position variances per particle
along the $x$ and $y$ directions,
$\frac{1}{N}\Delta_{\hat X}^2$ and $\frac{1}{N}\Delta_{\hat Y}^2$,
for $N=10^2,\ldots,10^6$ bosons as a function of the barrier height.
The interaction parameter is $\Lambda=\lambda_0(N-1)=0.1$.
The anisotropies of the many-particle position variances
at the many-body and mean-field levels of theory 
become opposite to each other at a barrier height just above $V_0=14$.
See the text for further details.
Actual data is marked by symbols, the continuous curves are to guide the eye only.
The quantities shown are dimensionless.
}
\label{f4}
\end{figure}

Fig.~\ref{f3} presents the depletion per particle, $1-\frac{n_1}{N}=\frac{\sum_{j>1} n_j}{N}$.
Here, $n_1$ is the largest occupation number 
obtained from diagonalization of the reduced one-particle 
density matrix $\rho^{(1)}(\r,\r')$.
Equivalently, the sum of all other occupation numbers, $\sum_{j>1} n_j$, 
is the total number of depleted particles.
We see that the total number of depleted particles increases with the barrier height $V_0$.
There are around $20$ bosons outside the condensed mode for $V_0=16$.
Whereas the total number of depleted particles saturates for increasing $N$ and a given barrier height,
the depletion per particle decreases.
This is in accordance with the first relation in (\ref{E2}).
To verify that $M=2$ self-consistent orbitals accurately describe the depletion
in the two-dimensional double well for all $V_0$,
we have performed calculations using $M=4$ and $N=100$ bosons,
which are seen to lie on top of the $M=2$ and $N=100$ bosons results,
see Fig.~\ref{f3}.
Since the interaction parameter is held fixed,
this implies the accuracy of the $M=2$ and $N>100$ results as well.
Furthermore,
we have found for $N=100$ bosons
that the system becomes two-fold fragmented with increasing barrier height 
\cite{MCHB,DWF1,DWF2,DWF3},
with $\frac{n_3}{N}, \frac{n_4}{N} < 10^{-7}$ for all $V_0$.
The two macroscopically-occupied natural orbitals are gerade and ungerade along the $x$ direction
and gerade along the $y$ direction.
The next two marginally-occupied natural orbitals are (for $V_0 \ge 8$)
ungerade along the $y$ direction
and gerade and ungerade along the $x$ direction. 
As the number of bosons is enlarged for a given barrier height
and the fixed interaction strength,
Fig.~\ref{f3} shows that the fragmentation disappears \cite{DWF2}
until $100\%$ condensation is obtained at the infinite-particle limit \cite{INF2}. 

Figs.~\ref{f1}, \ref{f2}, \ref{f3} have demonstrated numerically
the literature results on the density per particle, energy per particle, and depletion per particle
of a trapped bosonic system in two spatial dimensions
with a constant interaction parameter
at the infinite-particle limit 
\cite{INF1,INF2,INF3}.
We now move to investigate the many-particle position variances in the system.
Complementary results on the many-particle momentum variances and
the respective uncertainty products are given and briefly discussed in the Appendix. 
Fig.~\ref{f4} depicts the results.
It is instrumental to analyze them in view of
the study of bosons in the one-dimensional double well as a function of the barrier height
at the infinite-particle limit in \cite{VAR1}.
We begin with the variance per particle along the $x$ direction, $\frac{1}{N}\Delta_{\hat X}^2$,
the direction along which the barrier is ramped up.
The variance computed at the mean-field level of theory increases monotonously,
signifying the spread of the density seen in Fig.~\ref{f1}.
In contrast, the quantity 
computed and the many-body level 
reaches a maximum at about $V_0 \approx 9$,
then it starts to decrease.
There are about $0.04$ particles depleted then, see Fig.~\ref{f3}.
This is sufficient for the difference between the mean-field and many-body variances to
become qualitative.
With increasing barrier height, the many-body $\frac{1}{N}\Delta_{\hat X}^2$ decreases further.
The same qualitative difference is found in the one-dimensional double well \cite{VAR1}.

The variance per particle along the $y$ direction, $\frac{1}{N}\Delta_{\hat Y}^2$,
behaves completely different, see Fig.~\ref{f4}.
First, the variance is practically independent of the height of the barrier.
Second, its value is essentially $\frac{1}{\sqrt{2}}$, i.e., 
half the inverse frequency of the harmonic confinement along the $y$ direction.
This implies a very weak coupling between the $x$ and $y$ directions
of the interacting bosons in the studied two-dimensional double-well trap for all barrier heights,
at least as far as the transverse variance is examined.
We emphasize that no assumptions have been made on the shape of the orbitals
and many-boson wavefunction which are determined numerically self-consistently according to the variational principle.
For the respective essential independence of the momentum variance
along the $y$ direction on the barrier height, see the Appendix.
Third and finally,
the mean-field and many-body variances practically coincide.

We can now combine the results for the position variances along the $x$ and $y$ directions together.
At the mean-field level, $\frac{1}{N}\Delta_{\hat X}^2 > \frac{1}{N}\Delta_{\hat Y}^2$
for all barrier heights.
This is in line with Fig.~\ref{f1},
showing that the system's density is wider along the $x$ direction than along the $y$ direction.
At the many-body level,
$\frac{1}{N}\Delta_{\hat X}^2 > \frac{1}{N}\Delta_{\hat Y}^2$ up to barrier height of $V_0 \approx 14$,
see Fig.~\ref{f4}.
Just above this barrier height the situation reverses, namely,
$\frac{1}{N}\Delta_{\hat X}^2 < \frac{1}{N}\Delta_{\hat Y}^2$.
Thus, despite the fact that the density along the $x$ direction is wider than along the $y$ direction,
the many-particle position variance is smaller along the $x$ direction than along the $y$ direction.
This inverse relation between the density and position variance 
constitutes an opposite anisotropy of the many-particle position variance.
Furthermore,
the anisotropy $\frac{1}{N}\Delta_{\hat X}^2 < \frac{1}{N}\Delta_{\hat Y}^2$ of the ground state 
saturates when increasing the number of particles while keeping the interaction parameter
fixed, see Fig.~\ref{f4},
i.e., it persists in the infinite-particle limit.

The system considered here is a two-dimensional double-well potential with a barrier of height $V_0$.
The $x$ and $y$ directions are very weakly coupled,
at least at as far as the position variance is examined.
Just above $V_0=14$, when the position variance becomes anisotropic,
the total number of depleted particles is $\sum_{j>1} n_j < 10$.
It is intriguing that already less then $10$ bosons can have such a sizable effect 
on a trapped Bose-Einstein condensate at the infinite-particle limit which is $100\%$ condensed.
This makes investigating the anisotropy of the position variance in traps 
of other shapes a fundamental and appealing direction to follow.

\section{Summary and outlook}\label{Sum}

The {\it ground state} of a Bose-Einstein condensate in a two-dimensional trap potential
is analyzed numerically at the infinite-particle limit.
It is demonstrated that the anisotropy of the many-particle 
position variance can be opposite when computed
at the many-body and mean-field levels of theory,
despite the system being $100\%$ condensed.
It would be interesting to find analogous situations 
for the momentum variance and, in three spatial dimensions,
for the variance of the many-particle angular-momentum operator.

\ack

This research was supported by the Israel Science Foundation (Grant No.~600/15). 
We thank Raphael Beinke, Sudip Haldar, and Kaspar Sakmann for discussions.
Computation time on the BwForCluster and the
Cray XC40 system Hazelhen at the High Performance Computing Center
Stuttgart (HLRS) is gratefully acknowledged.

\appendix\section{Many-particle momentum variance and uncertainty product}\label{APP}

\begin{figure}[!]
\begin{center}
\includegraphics[width=0.545\columnwidth,angle=-90]{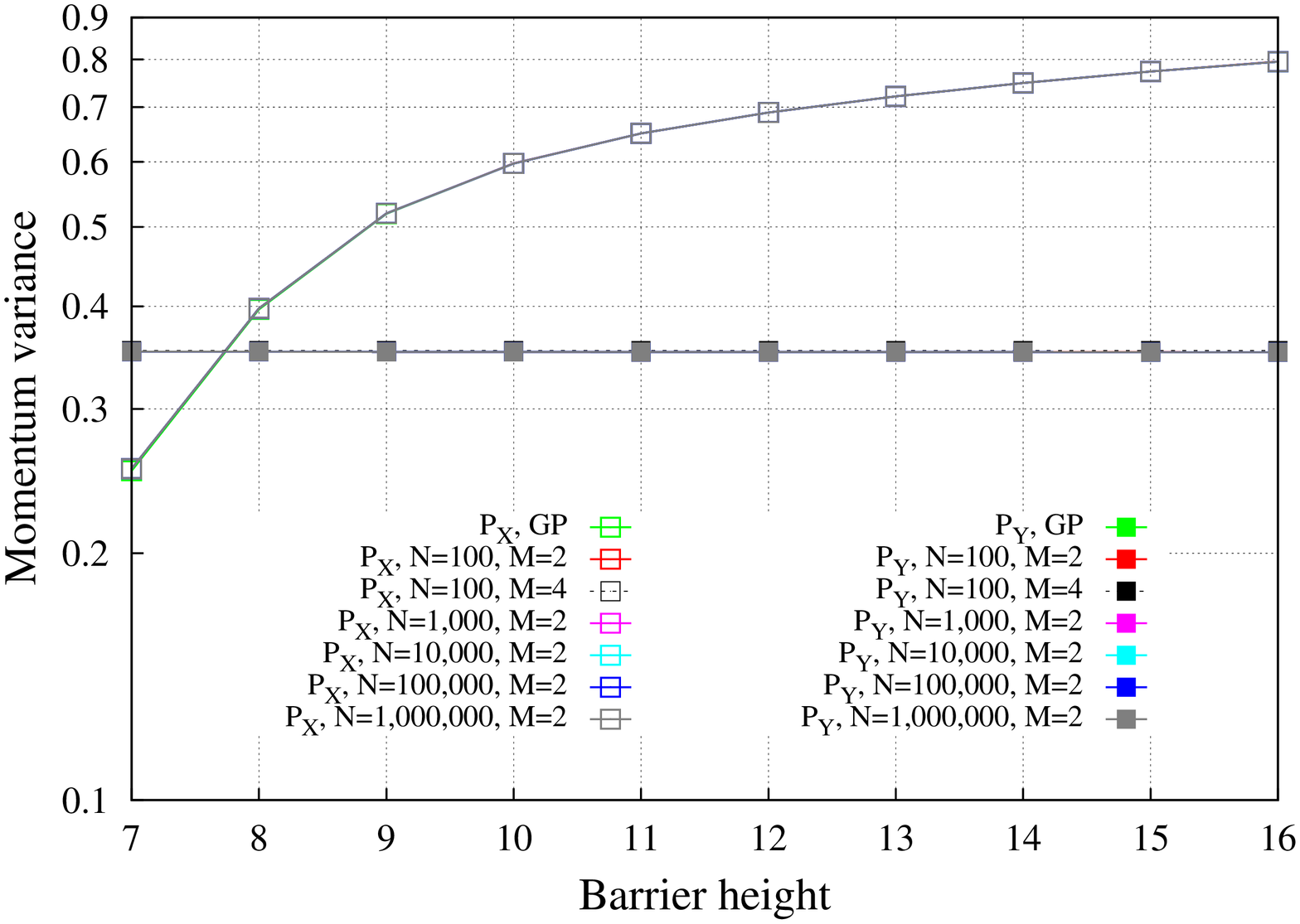}
\end{center}
\vglue 0.5 truecm
\caption{Many-particle momentum variances per particle
along the $x$ and $y$ directions,
$\frac{1}{N}\Delta_{\hat P_X}^2$ and $\frac{1}{N}\Delta_{\hat P_Y}^2$,
for $N=10^2,\ldots,10^6$ bosons as a function of the barrier height.
The interaction parameter is $\Lambda=\lambda_0(N-1)=0.1$.
The many-body and mean-field momentum quantities are essentially the same along
both the $x$ and $y$ directions.
See the text for further details.
Actual data is marked by symbols, the continuous curves are to guide the eye only.
The quantities shown are dimensionless.
}
\label{f5}
\end{figure}

The appendix presents complementary results.
Fig.~\ref{f5} presents the many-particle momentum variances per particle
$\frac{1}{N}\Delta_{\hat P_X}^2$ and $\frac{1}{N}\Delta_{\hat P_Y}^2$ as a function of the barrier height.
Unlike the position variance, here the respective many-body and mean-field quantities along the $x$
and the $y$ directions are essentially the same.
This means that there is no anisotropy
of the momentum variance for the ground state in the double-well potential,
let alone at the infinite-particle limit.
We remark that anisotropy of the momentum variance
has been found in the {\it out-of-equilibrium} scenario of \cite{VAR3}
at the infinite-particle limit.
It would be interesting to find a ground state
whose momentum variance exhibits
anisotropy at the infinite-particle limit.

The momentum variance along the $x$ direction grows monotonously with $V_0$.
This growth
of the momentum variance originates
from the narrowing of each of the two density peeks in their potential wells with increasing barrier height,
despite the overall broadening of the density along the $x$ direction, see Fig.~\ref{f1}.
The momentum variance along the $y$ direction is practically independent of the barrier height,
its value being essentially $\frac{1}{2\sqrt{2}}$, i.e.,
half the frequency of the harmonic trap along the $y$ direction.
This matches the finding for the position variance, see Fig.~\ref{f4},
and corroborates the very weak coupling between the $x$ and $y$ directions
found at the level of the many-particle position variance also at the
level of the many-particle momentum variance.

\begin{figure}[!]
\begin{center}
\includegraphics[width=0.545\columnwidth,angle=-90]{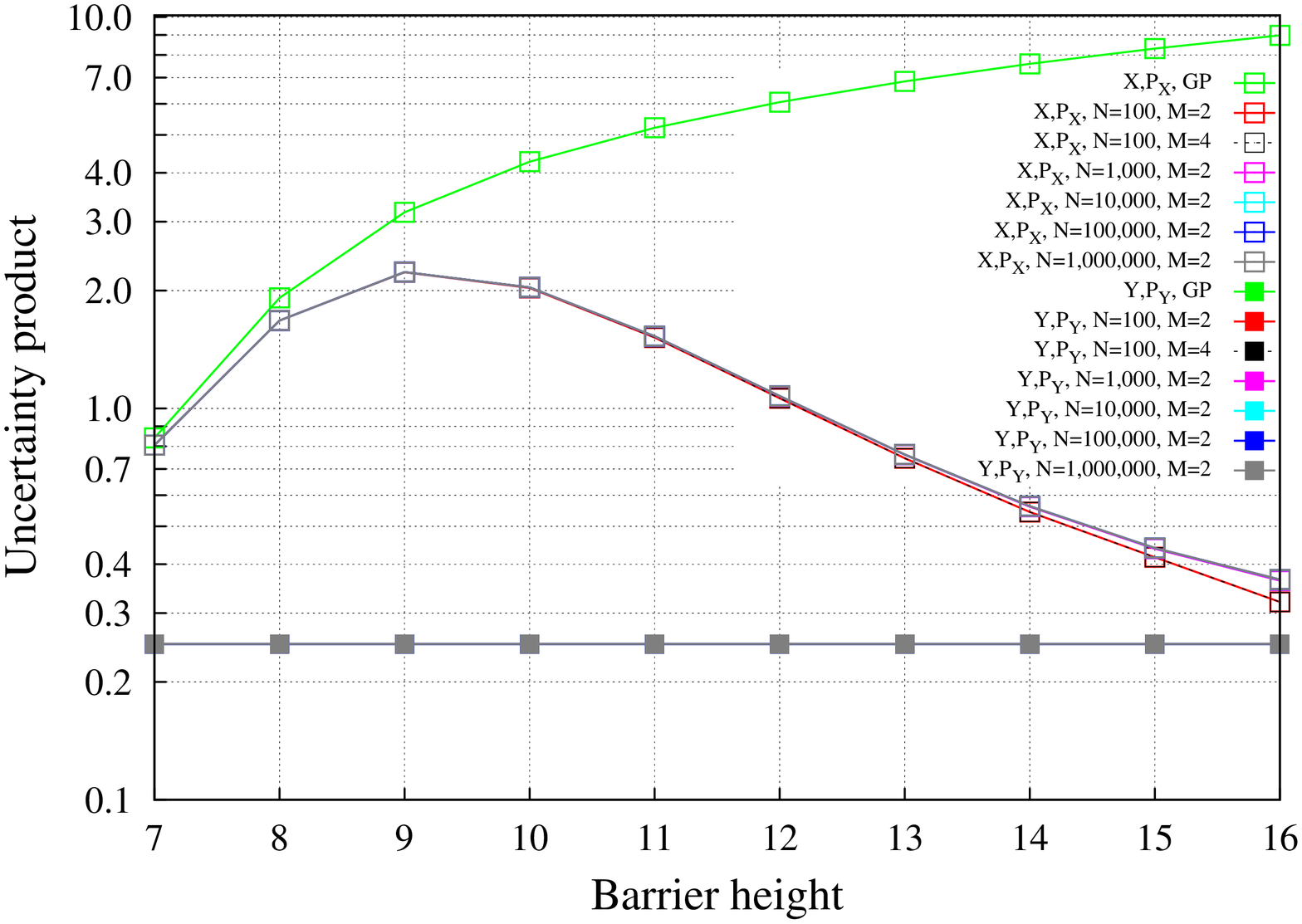}
\end{center}
\vglue 0.5 truecm
\caption{Many-particle uncertainty products
along the $x$ and $y$ directions,
$\frac{1}{N}\Delta_{\hat X}^2\frac{1}{N}\Delta_{\hat P_X}^2 = 
\Delta_{\hat X_{CM}}^2\Delta_{\hat P_{X_{CM}}}^2$ 
and $\frac{1}{N}\Delta_{\hat Y}^2\frac{1}{N}\Delta_{\hat P_Y}^2=
\Delta_{\hat Y_{CM}}^2\Delta_{\hat P_{Y_{CM}}}^2$,
for $N=10^2,\ldots,10^6$ bosons as a function of the barrier height.
The interaction parameter is $\Lambda=\lambda_0(N-1)=0.1$.
The many-body and mean-field uncertainty products differ substantially along the $x$ direction,
whereas they are essentially the same along the $y$ direction.
See the text for further details.
Actual data is marked by symbols, the continuous curves are to guide the eye only.
The quantities shown are dimensionless.
}
\label{f6}
\end{figure}

Last but not least,
we combine the results
for the position and momentum variances together.
This is most naturally done in terms of their uncertainty products,
$\frac{1}{N}\Delta_{\hat X}^2\frac{1}{N}\Delta_{\hat P_X}^2 = 
\Delta_{\hat X_{CM}}^2\Delta_{\hat P_{X_{CM}}}^2$ 
and $\frac{1}{N}\Delta_{\hat Y}^2\frac{1}{N}\Delta_{\hat P_Y}^2=
\Delta_{\hat Y_{CM}}^2\Delta_{\hat P_{Y_{CM}}}^2$ \cite{VAR1}.
Fig.~\ref{f6} presents the uncertainty products along the $x$ and $y$ directions.
Summing up all the above results,
we can see for the ground state at the infinite-particle limit that:
(i) The many-particle uncertainty product along the $x$ direction computed at the mean-field level of theory 
increases monotonously with the barrier height $V_0$,
whereas that computed at the many-body level increases first and then decreases.
(ii) The many-particle uncertainty products computed at the many-body and mean-field levels
along the $y$ direction are essentially independent of the barrier height,
and are practically minimal, $\frac{1}{\sqrt{2}} \times \frac{1}{2\sqrt{2}} = \frac{1}{4}$,
reflecting the very weak coupling of the $x$ and $y$ directions discussed above.
(iii) In the `competition' between the many-particle position variance,
which exhibits anisotropy, see Fig.~\ref{f4},
and the many-particle momentum variance which does not, see Fig.~\ref{f5},
the latter `wins';
$\frac{1}{N}\Delta_{\hat X}^2\frac{1}{N}\Delta_{\hat P_X}^2 >
\frac{1}{N}\Delta_{\hat Y}^2\frac{1}{N}\Delta_{\hat P_Y}^2$
for all considered
barrier heights $V_0$
at the infinite-particle limit.
This is a good place to conclude
our investigations.

\end{document}